\documentclass[aps,prd,twocolumn,superscriptaddress,showpacs]{revtex4}
\usepackage[dvips]{graphicx}
\usepackage{amsmath,amssymb,times}

\newcommand{\bequ}{\begin{equation}}
\newcommand{\eequ}{\end{equation}}
\newcommand{\bea}{\begin{eqnarray}}
\newcommand{\eea}{\end{eqnarray}}

\renewcommand{\a}{\alpha}

\DeclareSymbolFont{boldletters}{OML}{cmm} {b}{it}
\DeclareSymbolFontAlphabet{\mathbit}{boldletters}
\DeclareMathSymbol{\alpha}{\mathalpha}{letters}{"0B}
\DeclareMathSymbol{\beta}{\mathalpha}{letters}{"0C}
\DeclareMathSymbol{\gamma}{\mathalpha}{letters}{"0D}
\DeclareMathSymbol{\delta}{\mathalpha}{letters}{"0E}
\DeclareMathSymbol{\epsilon}{\mathalpha}{letters}{"0F}
\DeclareMathSymbol{\zeta}{\mathalpha}{letters}{"10}
\DeclareMathSymbol{\eta}{\mathalpha}{letters}{"11}
\DeclareMathSymbol{\theta}{\mathalpha}{letters}{"12}
\DeclareMathSymbol{\iota}{\mathalpha}{letters}{"13}
\DeclareMathSymbol{\kappa}{\mathalpha}{letters}{"14}
\DeclareMathSymbol{\lambda}{\mathalpha}{letters}{"15}
\DeclareMathSymbol{\mu}{\mathalpha}{letters}{"16}
\DeclareMathSymbol{\nu}{\mathalpha}{letters}{"17}
\DeclareMathSymbol{\xi}{\mathalpha}{letters}{"18}
\DeclareMathSymbol{\pi}{\mathalpha}{letters}{"19}
\DeclareMathSymbol{\rho}{\mathalpha}{letters}{"1A}
\DeclareMathSymbol{\sigma}{\mathalpha}{letters}{"1B}
\DeclareMathSymbol{\tau}{\mathalpha}{letters}{"1C}
\DeclareMathSymbol{\upsilon}{\mathalpha}{letters}{"1D}
\DeclareMathSymbol{\phi}{\mathalpha}{letters}{"1E}
\DeclareMathSymbol{\chi}{\mathalpha}{letters}{"1F}
\DeclareMathSymbol{\psi}{\mathalpha}{letters}{"20}
\DeclareMathSymbol{\omega}{\mathalpha}{letters}{"21}
\DeclareMathSymbol{\varepsilon}{\mathalpha}{letters}{"22}
\DeclareMathSymbol{\vartheta}{\mathalpha}{letters}{"23}
\DeclareMathSymbol{\varpi}{\mathalpha}{letters}{"24}
\DeclareMathSymbol{\varrho}{\mathalpha}{letters}{"25}
\DeclareMathSymbol{\varsigma}{\mathalpha}{letters}{"26}
\DeclareMathSymbol{\varphi}{\mathalpha}{letters}{"27}
\DeclareMathSymbol{\Gamma}{\mathalpha}{letters}{"00}
\DeclareMathSymbol{\Delta}{\mathalpha}{letters}{"01}
\DeclareMathSymbol{\Theta}{\mathalpha}{letters}{"02}
\DeclareMathSymbol{\Lambda}{\mathalpha}{letters}{"03}
\DeclareMathSymbol{\Xi}{\mathalpha}{letters}{"04}
\DeclareMathSymbol{\Pi}{\mathalpha}{letters}{"05}
\DeclareMathSymbol{\Sigma}{\mathalpha}{letters}{"06}
\DeclareMathSymbol{\Upsilon}{\mathalpha}{letters}{"07}
\DeclareMathSymbol{\Phi}{\mathalpha}{letters}{"08}
\DeclareMathSymbol{\Psi}{\mathalpha}{letters}{"09}
\DeclareMathSymbol{\Omega}{\mathalpha}{letters}{"0A}

\begin{document}
\preprint{SAGA-HE-257}
\title{Equation of state in the PNJL model with the entanglement interaction}

\author{Yuji Sakai}
\email[]{sakai@phys.kyushu-u.ac.jp}
\affiliation{Department of Physics, Graduate School of Sciences, Kyushu University,
             Fukuoka 812-8581, Japan}

\author{Takahiro Sasaki}
\email[]{sasaki@phys.kyushu-u.ac.jp}
\affiliation{Department of Physics, Graduate School of Sciences, Kyushu University,
             Fukuoka 812-8581, Japan}

\author{Hiroaki Kouno}
\email[]{kounoh@cc.saga-u.ac.jp}
\affiliation{Department of Physics, Saga University,
             Saga 840-8502, Japan}

\author{Masanobu Yahiro}
\email[]{yahiro@phys.kyushu-u.ac.jp}
\affiliation{Department of Physics, Graduate School of Sciences, Kyushu University,
             Fukuoka 812-8581, Japan}

\date{\today}

\begin{abstract}
The equation of state and the phase diagram in two-flavor QCD are 
investigated by the Polyakov-loop extended Nambu--Jona-Lasinio (PNJL) model 
with an entanglement vertex between the chiral condensate and the 
Polyakov-loop. The entanglement-PNJL (EPNJL) model reproduces 
LQCD data at $\mu \ge 0$ better than the PNJL model. 
Hadronic degrees of freedom are taken into account by 
the free-hadron-gas (FHG) model with the volume-exclusion effect 
due to the hadron generation. 
The EPNJL+FHG model improves agreement of the EPNJL model with 
LQCD data particularly at small temperature. 
The quarkyonic phase survives, even if the correlation between 
the chiral condensate and the Polyakov loop is strong and  
hadron degrees of freedom are taken into account. 
However, the location of the quarkyonic phase is sensitive 
to the strength of the volume exclusion. 
\end{abstract}

\maketitle

\section{Introduction}
\label{Introduction}

A key issue in the thermodynamics of quantum chromodynamics (QCD) 
is whether the chiral-symmetry restoration and 
the confinement-to-deconfinement transition coincide or not. 
If the two transitions do not coincide, phases such as 
constituent quark phase~\cite{Cleymans,Kouno1} or 
quarkyonic phase~\cite{McLerran1,Fukushima2,Abuki,Sasaki-C} may appear. 
When the two transitions are first order, 
discontinuities can appear simultaneously 
in their (approximate) order parameters, 
chiral condensate $\sigma$ and Polyakov loop $\Phi$~\cite{BCGG,Kashiwa5}. 
At zero quark chemical potential ($\mu$), however, 
the two transitions are found to be crossover 
by lattice QCD (LQCD)~\cite{Karsh,S-Aoki,Y-Aoki}. 
Hence, there is no a priori reason why the two transitions coincide. 
Actually, there is a debate~\cite{Borsanyi} 
on the coincidence in LQCD data 
at zero $\mu$~\cite{Karsh,S-Aoki,Y-Aoki}.  

LQCD has the sign problem for finite $\mu$~\cite{Kogut1}. 
This is an important problem to be solved in future. 
Fortunately, LQCD data are available at imaginary $\mu$
~\cite{FP,Chen,D'Elia-RW,FP-RW,Nagata,Takaishi,Lombardo} 
and real and imaginary isospin chemical potential~\cite{Kogut2,Cea, D'Elia-Iso}  because of no sign problem in the regions. The data show that chiral and 
deconfinement crossovers coincide within the numerical accuracy. 
This coincidence indicates that 
there exists a strong correlation (entanglement) 
between $\sigma$ and $\Phi$. 

An approach complementary to first-principle LQCD with the sign problem 
is to build an 
effective model consistent with LQCD data and apply the model to the 
real-$\mu$ region. 
The Polyakov-loop extended Nambu--Jona-Lasinio (PNJL) 
model is designed to treat both 
the chiral-symmetry restoration and the deconfinement transition
~\cite{Meisinger,Fukushima,Ratti,Rossner,Schaefer,Kouno,Sakai1,Sakai2,Sakai3,Sakai4,Sakai5,Sasaki,Kashiwa2,Kashiwa3,Matsumoto}. 
For the thermal system with temperature ($T$) 
and imaginary chemical potential $\mu=i\theta T$,
LQCD has the periodicity of $2\pi/3$ in $\theta$. 
For temperatures higher than some critical temperature $T_{\rm E}$, 
LQCD has the first-order phase transition 
at $\theta=\pi/3$ mod $2\pi/3$. 
These periodicity and transition were first proposed by 
Roberge and Weiss~\cite{RW}, and then called 
the RW periodicity and the RW phase transition, 
respectively. 
The PNJL model can reproduce LQCD data for 
the RW periodicity and the RW phase 
transition~\cite{Sakai1,Sakai2,Sakai3,Sakai4,Sakai5,Kouno}, 
but not for the coincidence between chiral and deconfinement 
crossovers~\cite{Sakai2}. 
A current topic at imaginary $\mu$ is the 
order of the RW transition at the endpoint $T=T_{\rm E}$. 
Recent LQCD shows that 
the order is first-order for small and large quark masses, but 
the order is weakened and could be second order 
at intermediate masses~\cite{D'Elia-RW,FP-RW}. 
The PNJL model cannot reproduce the property~\cite{Sakai-E}.

In order to solve these problems in the PNJL model, 
we recently extended the model so as to have 
a four-quark vertex depending on $\Phi$. 
If the gluon field $A_{\nu}$ has a vacuum expectation value 
$\langle A_{0} \rangle$ in its time component, 
such a four-quark vertex depending on $\Phi$ 
is allowed through $\langle A_{0} \rangle$~\cite{Kondo}. 
The vertex introduced newly induces a strong correlation (entanglement) 
between $\sigma$ and $\Phi$. The strong correlation is 
supported by recent calculations~\cite{Braun} of 
the exact renormalization-group equation~\cite{Wetterich}.  
The new model is called the entanglement-PNJL (EPNJL) model~\cite{Sakai-E}.
The EPNJL results are consistent with LQCD data 
at zero and imaginary $\mu$ and also at finite isospin-chemical 
potential~\cite{Sakai-E}. 
Particularly, the model can reproduce not only the coincidence 
between chiral and deconfinement crossovers 
without~\cite{Sakai-E} and 
with the strong magnet field~\cite{Gatto:2010pt} 
but also the quark-mass dependence of the order of 
the RW endpoint mentioned above~\cite{Sakai-E}. 

Another current topic related to the correlation between 
the chiral and deconfinement transitions is 
the quarkyonic phase~\cite{McLerran1,Fukushima2,Abuki,Sasaki-C}. 
The concept of the phase is originally constructed 
in the limit of a large number of colors ($N_{\rm c}$). 
It is a confined phase 
with finite quark-number density ($n_{\rm q}$). 
Since the $n_{\rm q}$-generation is strongly induced 
by the chiral restoration, 
the quarkyonic phase is nearly equal to the chirally-symmetric 
confined phase. 
Recently, the PNJL calculations showed that 
the quarkyonic phase survives even 
for small $N_{\rm c}$ such as 3~\cite{McLerran1,Fukushima2,Abuki,Sasaki-C}. 
It is then interesting whether the phase can survive in the 
EPNJL model with the strong correlation between $\sigma$ and $\Phi$. 
Another point to be argued is the relation between the quarkyonic phase and 
hadron degrees of freedom, since the location of the phase was always 
discussed without thinking the degrees of freedom for the case of 
$N_{\rm c}=3$.

In this paper, we analyze the thermodynamics of two-flavor QCD 
with the EPNJL model and its simple extension. 
First, we test the reliability of the EPNJL model by comparing the model 
results with LQCD data on the equation of state (EOS) at $\mu \ge 0$. 
After confirming the reliability of the EPNJL model, 
we predict the phase diagram in the $\mu$-$T$ plane and investigate 
whether the quarkyonic phase survives under the strong correlation between 
$\sigma$ and $\Phi$. Hadronic degrees of freedom are introduced 
by the free-hadron-gas (FHG) model with constant or 
$T$- and $\mu$-dependent hadron masses. Here, the volume exclusion effect 
due to the hadron generation~\cite{Rischke,Steinheimer}~is also taken into account in the FHG model. 
The volume exclusion effect guarantees the quark dominance 
in EOS at high $T$ and/or high $\mu$. 
EOS, the phase diagram and the presence or absence of the quarkyonic phase 
are investigated also by the EPNJL+FHG model. 

 In Section \ref{PNJL}, the PNJL model is recapitulated. 
 In Section \ref{Results}, the FHG model is explained briefly and 
 numerical results of the EPNJL and EPNJL+FHG models are shown. 
 Section \ref{Summary} is devoted to summary.

\section{PNJL model}
\label{PNJL}

We start with the standard two-flavor PNJL Lagrangian~\cite{Fukushima2,Fukushima,Ratti}
\begin{align}
 {\cal L}  =& {\bar q}(i \gamma_\nu D^\nu -m_0)q \notag\\
             &\hspace{3mm} + G_{\rm s}[({\bar q}q)^2 
                          +({\bar q}i\gamma_5 {\vec \tau}q)^2] 
              - {\cal U}(\Phi [A],{\Phi} [A]^*,T) ,
             \label{eq:E1}
\end{align}
where $q$ denotes the two-flavor quark field, 
$m_0$ represents the current quark mass, 
and $D^\nu=\partial^\nu+iA^\nu-i\mu\delta^{\nu}_{0}$. 
Field $A^\nu$ is defined by $A^\nu=\delta^{\nu}_{0}gA^0_a{\lambda^a\over{2}}$ 
with gauge fields $A^\nu_a$, the Gell-Mann matrix $\lambda_a$, and the 
gauge coupling $g$.
In the NJL sector, $G_{\rm s}$ denotes the coupling constant of 
the scalar-type four-quark interaction. 
The Polyakov potential ${\cal U}$, defined in (\ref{eq:E13}), is a function 
of Polyakov loop $\Phi$ and its Hermitian conjugate $\Phi^*$,
\begin{align}
\Phi      = {1\over{N_{\rm c}}}{\rm Tr} L,~~~~
\Phi^{*}  = {1\over{N_{\rm c}}} {\rm Tr}L^\dag ,
\end{align}
with
\begin{align}
L({\bf x}) = {\cal P} \exp\Bigl[
                {i\int^\beta_0 d \tau A_4({\bf x},\tau)}\Bigr],
\end{align}
where ${\cal P}$ is the path ordering and $A_4 = iA_0 $. 
In the chiral limit ($m_0=0$), the Lagrangian density has the exact $SU(N_f)_{\rm L} \times SU(N_f)_{\rm R}\times U(1)_{\rm v} \times SU(3)_{\rm c}$  symmetry. 
The temporal component of the gauge field is diagonal in flavor space, 
because color and flavor spaces are completely separated 
in the present case. 
In the Polyakov gauge, $L$ can be written in a diagonal form 
in color space~\cite{Fukushima}: 
\begin{align}
L 
=  e^{i \beta (\phi_3 \lambda_3 + \phi_8 \lambda_8)}
= {\rm diag} (e^{i \beta \phi_a},e^{i \beta \phi_b},
e^{i \beta \phi_c} ),
\label{eq:E6}
\end{align}
where $\phi_a=\phi_3+\phi_8/\sqrt{3}$, $\phi_b=-\phi_3+\phi_8/\sqrt{3}$
and $\phi_c=-(\phi_a+\phi_b)=-2\phi_8/\sqrt{3}$. 
The Polyakov loop $\Phi$ is an exact order parameter of spontaneous 
${\mathbb Z}_3$ symmetry breaking in pure gauge theory.
Although ${\mathbb Z}_3$ symmetry is not an exact one 
in the system with dynamical quarks, it still seems to be a good indicator of 
the deconfinement phase transition. 
Therefore, we use $\Phi$ to define the deconfinement phase transition.

Making the mean field approximation and performing 
the path integral over the quark field, 
one can obtain the thermodynamic potential $\Omega$ (per volume), 
\begin{align}
&\Omega = -2 N_f\int \frac{d^3{\rm p}}{(2\pi)^3}
      	\left[ 3 E ({\bf p})
	 + \frac{1}{\beta}\ln~ {\cal F}_q {\cal F}_{\bar{q}}\right]
 	 + U_{\rm M}+{\cal U}, 
\label{eq:E12} \\
&~~~{\cal F}_q = 1 + 3(\Phi+\Phi^* e^{-\beta E^-})e^{-\beta E^-}
 + e^{-3\beta E^-}\notag\\
&~~~{\cal F}_{\bar q} = 1 + 3(\Phi^*+\Phi e^{-\beta E^+})e^{-\beta E^+}
 + e^{-3\beta E^+}
\label{Omega}
\end{align}
where $\sigma = \langle \bar{q}q \rangle $, $M=m_0 -2 G_{\rm s} \sigma$, 
$U_{\rm M}= G_{\rm s} \sigma^2$, $E=\sqrt{{\bf p}^2+M^2}$ and 
$E^\pm=E\pm \mu$. On the right-hand side of \eqref{eq:E12}, 
only the first term  diverges. 
It is then regularized by the three-dimensional momentum
cutoff $\Lambda$~\cite{Fukushima,Fukushima2,Ratti}.
The variables $X=\Phi$, ${\Phi}^*$ and $\sigma$ 
satisfy the stationary conditions $\partial \Omega/\partial X=0$.

We use ${\cal U}$ of Ref.~\cite{Rossner}, which is fitted to LQCD 
data in pure gauge theory at finite $T$~\cite{Boyd,Kaczmarek}: 
\begin{align}
&{\cal U} = T^4 \Bigl[-\frac{a(T)}{2} {\Phi}^*\Phi\notag\\
      &~~~~~+ b(T)\ln(1 - 6{\Phi\Phi^*}  + 4(\Phi^3+{\Phi^*}^3)
            - 3(\Phi\Phi^*)^2 )\Bigr] 
            \label{eq:E13}
\end{align}
with 
\begin{align}
a(T)   = a_0 + a_1\Bigl(\frac{T_0}{T}\Bigr)
                 + a_2\Bigl(\frac{T_0}{T}\Bigr)^2,~~~~
b(T)=b_3\Bigl(\frac{T_0}{T}\Bigr)^3  , 
\label{eq:E14}
\end{align}
where the parameters are summarized in Table~\ref{Table1}. 
In pure gauge theory, the Polyakov potential yields a first-order 
deconfinement phase transition at $T=T_0$.
The original value of $T_0$ is 270 MeV that reproduces pure gauge LQCD 
data, but the PNJL model with the value of $T_0$ yields a larger value of 
the pseudocritical temperature $T_{\rm c}$ at zero chemical potential than 
$T_{\rm c}=173\pm 8$~MeV given by full LQCD 
data~\cite{KL1994,Karsch4,Kaczmarek2}.  
Therefore, we rescale $T_0$ to 212 MeV in the PNJL model so as to reproduce 
LQCD result, $T_{\Phi}=173$~MeV, for 
the deconfinement transition temperature. 
However, the PNJL calculation yields 
the chiral transition temperature $T_{\sigma}=216$~MeV, while 
LQCD gives $T_{\sigma}=173$~MeV. 
Therefore, the PNJL model has a sizable difference 
between $T_{\sigma}$ and $T_{\Phi}$, 
say $\Delta=|T_{\sigma}-T_{\Phi}|/T_{\Phi} \approx 20 \%$~\cite{Sakai2}. 
Thus, the PNJL result is not consistent 
with LQCD data for $T_{\sigma}$ and hence $\Delta$.

\begin{table}[h]
\begin{center}
\begin{tabular}{llllll}
\hline
~~~~~$a_0$~~~~~&~~~~~$a_1$~~~~~&~~~~~$a_2$~~~~~&~~~~~$b_3$~~~~~
\\
\hline
~~~~3.51 &~~~~-2.47 &~~~~15.2 &~~~~-1.75\\
\hline
\end{tabular}
\caption{
Summary of the parameter set in the Polyakov sector
used in Ref.~\cite{Rossner}. 
All parameters are dimensionless. 
\label{Table1}
}
\end{center}
\end{table}

The sizable difference indicates that the entanglement between the 
chiral and deconfinement transitions is weak in the PNJL model. 
In order to solve this problem, we proposed an effective coupling 
depending on the Polyakov loop, $G_{\rm s}(\Phi)$. 
In fact, this vertex is discussed in the exact renormalization group method
~\cite{Kondo}. 
It is expected that $\Phi$ dependence of $G_{\rm s}(\Phi)$ will be 
determined in future by an exact method such as the 
exact renormalization group 
method~\cite{Braun,Kondo,Wetterich}. 
In this paper, however, we simply assume the following $G_{\rm s}(\Phi)$ 
by respecting chiral symmetry, C symmetry~\cite{Kouno} and 
extended $\mathbb{Z}_3$ symmetry~\cite{Sakai-E}: 
\begin{eqnarray}
G_{\rm s}(\Phi)=G_{\rm s}[1-\alpha_1\Phi\Phi^*-\alpha_2(\Phi^3+\Phi^{*3})]. 
\label{entanglement interaction}
\end{eqnarray}
Thus, this model has entanglement interactions between $\sigma$ and $\Phi$ 
in addition to the covariant derivative in the original PNJL model.
The PNJL model with the entanglement vertex $G_{\rm s}(\Phi)$ is 
referred to as entanglement-PNJL (EPNJL) model. 
The parameters, $\alpha_1$ and $\alpha_2$, determined from 
LQCD data at zero and imaginary chemical potentials 
are $\alpha_1=\alpha_2=0.2$~\cite{Sakai-E}. 
Hadron degrees of freedom will be considered in 
Section~\ref{Results}.

\section{Results}
\label{Results}

\subsection{Comparison of model calculation to LQCD data at $\mu \ge 0$}

\begin{figure}[htbp]
\begin{center}
 \includegraphics[width=0.35\textwidth]{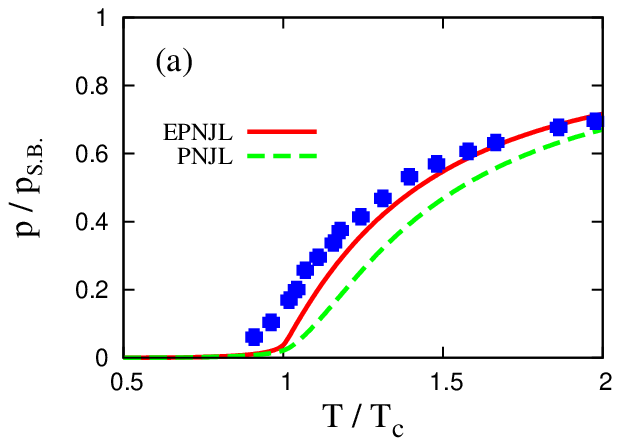} 
 \includegraphics[width=0.35\textwidth]{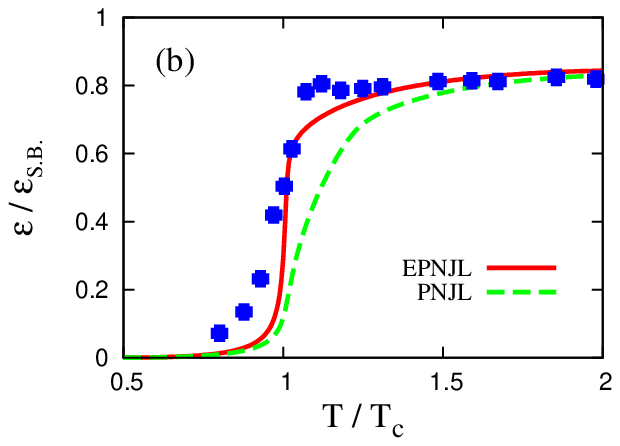} 
 \includegraphics[width=0.35\textwidth]{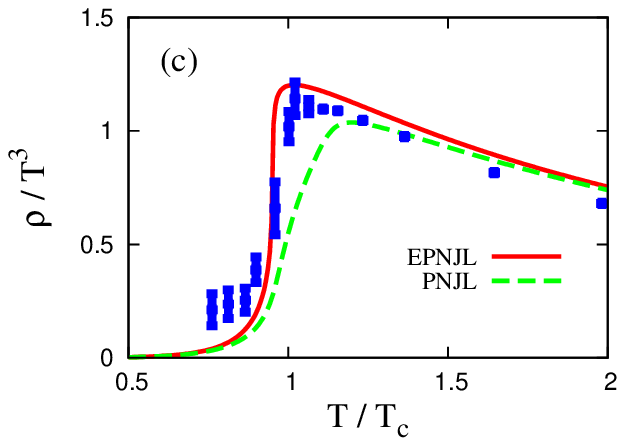} 
\end{center}
\caption{$T$ dependence of (a) the pressure at $\mu=0$, 
(b) the energy density at $\mu=0$ and (c) the net quark-number density 
at $\mu/T_{\rm c}=0.8$. 
The pressure and the energy density are divided by the values 
in the Stefan-Boltzmann limit at $\mu=0$, while the net quark-number 
density is 
divided by $T^3$. 
The solid (dashed) lines show the EPNJL (PNJL) result. 
LQCD data are taken from Ref.~\cite{Khan} for panels (a) and (b) and 
Ref.~\cite{Allton} for panel (c). 
}
\label{PEN1}
\end{figure}

The equations of state (EOS) calculated with 
the PNJL and EPNJL models are compared with LQCD data 
at $\mu \ge 0$~\cite{Khan,Allton}. 
Figure~\ref{PEN1} shows the pressure $p(T)$ and 
the energy density $\varepsilon(T)$ at $\mu=0$ and 
the net quark-number density $\rho(T)$ at $\mu=0.8T_{\rm c}$, 
where $\rho=n_{\rm q}-{\bar n}_{\rm q}$ for 
the quark number density $n_{\rm q}$ and 
the antiquark number density ${\bar n}_{\rm q}$ and 
$T_{\rm c}$ is the pseudocritical temperature of 
the deconfinement transition at $\mu=0$, that is, the peak position 
of the Polyakov-loop susceptibility. 
For $p(T)$ and $\varepsilon(T)$, 
LQCD data~\cite{Khan} provide 
only the deviations, $p(T)-p(T_{\rm n})$ 
and $\varepsilon(T)-\varepsilon(T_{\rm n})$, 
from $T_{\rm n}=0.9T_{\rm c}$.
Hence, $p(T_{\rm n})$ and $\varepsilon(T_{\rm n})$ are evaluated by 
the free-gas model of hadrons with vacuum masses. 
This procedure is reliable at $T=0.9T_{\rm c}$, 
because $p(T)$ and $\varepsilon(T)$ are dominated 
by the hadron components there; we will discuss this point later in 
subsection~\ref{hadronic effect on phase diagram}. 
The $p(T)$ and $\varepsilon(T)$ thus estimated from the LQCD data 
are shown by the dots in panels (a) and (b); note that in these panels 
the $p(T)$ and $\varepsilon(T)$ are normalized 
by the values in the Stefan-Boltzmann limit at $\mu=0$. 
In panel (c), $\rho(T)$ is nondimensionalized by $T^{3}$.

For all of $p$, $\varepsilon$ and $\rho$, 
the EPNJL results (solid lines) are more consistent with 
LQCD data than the PNJL results (dashed lines), as shown in Fig.~\ref{PEN1}. 
The entanglement interaction makes the chiral 
symmetry restoration faster, that is, it 
shifts $T_{\sigma}$ down to $T_{\Phi}$~\cite{Sakai-E}. 
As a consequence of the shift-down property, the EPNJL model has 
rapid change in $p(T)$, $\varepsilon(T)$ and $\rho(T)$ with $T$ and hence 
reproduces the sharp change of LQCD result better than the PNJL model. 
Both the EPNJL and PNJL models underestimate LQCD results at $T < T_{\rm c}$, 
because these models have no hadron component 
in their $p(T)$, $\varepsilon(T)$ and $\rho(T)$. We will return this point 
later in subsection~\ref{hadronic effect on phase diagram}.

\subsection{Phase diagram in $T-\mu$ plane based on the EPNJL model}

The pseudocritical temperature $T_{\Phi}$ of the deconfinement crossover 
was defined in two ways so far. 
One uses the peak of the Polyakov-loop susceptibility 
$\chi_{\Phi}$~\cite{Abuki,Sasaki-C}
and another does $\Phi=0.5$~\cite{Fukushima2}. 
The two definitions yield almost the same value of $T_{\Phi}$ 
for lower $\mu$, but for higher $\mu$ the former becomes somewhat obscure 
since $\chi_{\Phi}$ has a broad peak there. 
We then take the latter definition in this paper.

\begin{figure}[htbp]
\begin{center}
 \includegraphics[width=0.35\textwidth]{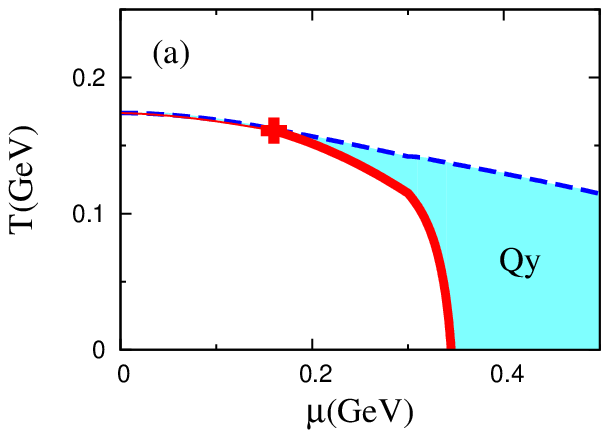} 
 \includegraphics[width=0.35\textwidth]{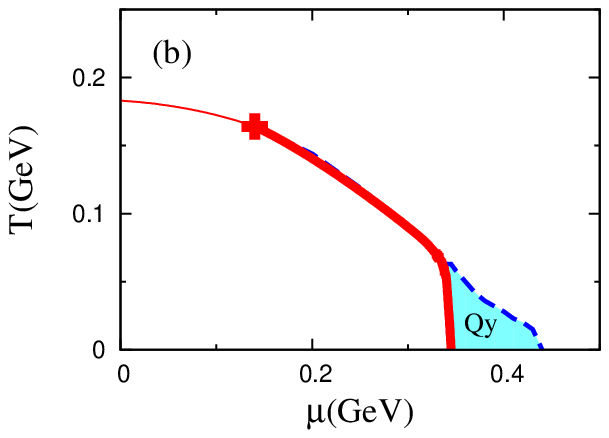} 
\end{center}
\caption{Phase diagram predicted by the EPNJL model with (a) 
constant $T_0$ and (b) $\mu$-dependent $T_0$. 
The thick (thin) solid line corresponds to the chiral transition line of 
first-order (crossover), while  
the dashed line represents the deconfinement crossover defined by $\Phi=0.5$. 
The plus symbol stands for the critical endpoint of the first-order 
chiral transition. 
The hatching region labeled by 'Qy' stands for the quarkyonic phase of 
definition 1. 
}
\label{PD}
\end{figure}

Figure~\ref{PD} (a) shows the phase diagram predicted by the EPNJL model with 
constant $T_0$. The thin-solid line represents the chiral crossover defined 
by the peak of the chiral susceptibility, while 
the dashed line shows the deconfinement crossover defined by $\Phi=0.5$. 
The thick-solid line stands for the first-order chiral transition. 
For $\mu \lesssim 0.15$~GeV, the entanglement interaction 
makes the chiral and deconfinement crossovers almost coincide. 
For $\mu \gtrsim 0.15$~GeV, however, the first-order chiral transition line
and the deconfinement crossover line diverge, 
so that there appears a chirally-symmetric but confined phase between 
the two lines. 
Since the $n_{\rm q}$-generation is strongly induced 
by the chiral restoration, 
the chiral transition line corresponds to the $n_{\rm q}$-generation line. 
In this sense, the chirally-symmetric confined phase is 
the quarkyonic phase in which quarks are confined but $n_{\rm q}$ is finite. 
Thus, the entanglement interaction does not make the two transitions coincide 
for $\mu \gtrsim 0.15$~GeV. This can be understood as below.

When the entanglement interaction is switched off, in general,  
$T_{\Phi}$ differs from $T_{\sigma}$. 
When $T_{\sigma} > T_{\Phi}$, $\Phi$ is large at $T_{\Phi}< T < T_{\sigma}$, 
so that the four-quark interaction \eqref{entanglement interaction} 
is suppressed there by the entanglement terms having $\a_1$ and $\a_2$. 
This suppression shifts 
$T_{\sigma}$ down to $T_{\Phi}$. 
This is the situation for $\mu \lesssim 0.15$~GeV. 
When $T_{\sigma} < T_{\Phi}$, meanwhile, $\Phi$ is small 
at $T_{\sigma}< T < T_{\Phi}$, 
so that little entanglement effect 
occurs in \eqref{entanglement interaction}. 
This is the case for $\mu \gtrsim 0.15$~GeV. 
Thus, the entanglement effect takes place mainly when $T_{\sigma} > T_{\Phi}$.

In principle, the Polyakov potential ${\cal U}$ may depend on $\mu$ as a 
consequence of the backreaction of the Fermion sector to the gluon sector. 
In fact, $\mu$ dependence of $T_0$ in ${\cal U}$ was estimated 
with the renormalization-group argument~\cite{Schaefer}:
\begin{eqnarray}
T_{0}(\mu)&=&T_{\tau}e^{-\frac{1}{\alpha_{0} b(\mu)}}
\end{eqnarray}
for $b(\mu)=29/(6\pi)-32\mu^2/(\pi T_{\tau}^2)$ with 
$\alpha_{0}=0.304$ and $T_{\tau}=1.770~{\rm [GeV]}$. 
Figure~\ref{PD}(b) shows an effect of $\mu$-dependent $T_{0}$ 
on the phase diagram. 
The chiral transition line is not changed much by 
introducing the $\mu$-dependent $T_0$. 
However, $T_{\Phi}$ lowers more for larger $\mu$, since so does $T_0$. 
Eventually, the chirally-symmetric confined phase, 
i.e., the quarkyonic phase, between the first-order chiral transition line 
and the deconfinement crossover line shrinks.

The concept of the quarkyonic phase~\cite{McLerran1} is originally constructed 
in the large $N_{\rm c}$ limit. It is a confined (color-singlet) phase 
with finite $n_{\rm q}$. 
The definition of the quarkyonic phase becomes somewhat unclear 
for small $N_{\rm c}=3$, since the deconfinement transition is 
crossover there. The confined state is defined by $\Phi < 0.5$, but 
the color-single state is by $\Phi=0$. 
Hence, a possible definition of the quarkyonic phase is 
a phase of $n_{\rm q} \neq 0$ and $\Phi < 0.5$. 
Another possible definition is a phase of $n_{\rm q} \neq 0$ 
and $\Phi =0$ that can exist only at $T=0$ and $\mu > M =340$~MeV. 
The second definition seems to be too strict, when the deconfinement 
transition is crossover. So we think the third possibility with the PNJL model. 
For this purpose, the net quark-number density $\rho$ is 
divided into three components 
\begin{align}
\rho=-\frac{\partial\Omega}{\partial\mu}=\rho_{1}+\rho_{2}+\rho_{3},
\label{n-decomposition} 
\end{align}
where 
\begin{align}
\rho_{k}=2N_f\int \frac{d^3{\rm p}}{(2\pi)^3}
\left[\frac{k~m_k({\bf p})}{1+\sum_j m_j({\bf p})}
-\frac{k~\bar{m}_k({\bf p})}{1+\sum_j\bar{m}_j({\bf p})}\right] 
\label{n-k} 
\end{align}
with 
\begin{align}
&~~m_1=3\Phi e^{-\beta{E_-}},~~\bar{m}_1=3\Phi^* e^{-\beta{E_+}},
~~m_2=3\Phi^* e^{-2\beta{E_-}},
\notag\\
&~~~~\bar{m}_2=3\Phi e^{-2\beta{E_+}},~~m_3=e^{-3\beta{E_-}},
~~\bar{m}_3=e^{-3\beta{E_+}}.
\notag
\end{align}
Here, $\rho_{k}$ means the net quark-number density where $k$ quarks or 
$k$ antiquarks are 
statistically in the same state~\cite{Fukushima2,Abuki}. 
The first (second) term on the right-hand side of \eqref{n-k} 
represents the quark (antiquark) contribution. 
In order for a phase to be color-singlet, three quarks must have 
the same momentum. 
This indicates that the quarkyonic phase can be defined by 
the $\rho_{3}$-dominated region of $\rho_{3} > \rho_{1}, \rho_{2}$~\cite{Fukushima2,Abuki}. 
This statement is reliable as follows. 
For zero temperature, $\rho$ is finite only when $\mu>M$, 
that is, at $\mu > 0.34$~GeV.
In the region of small $T$ and $\mu > 0.34$~GeV, furthermore, 
$\rho$ is dominated by $\rho_{3}$ because of $e^{\beta\mu} \gg 1$ 
in \eqref{n-decomposition}. 
At $\mu > M$ and low $T$ where the $\rho_{3}$-dominated region emerges, 
the quark part  is larger than the antiquark part in \eqref{n-k}, 
and the denominator ${1+\sum_j m_j({\bf p})}$ of the quark part is 
dominated by ${\mathbb Z}_{3}$-invariant $m_3$. 
Thus, the $\rho_3$-dominated region possesses 
the color-singlet nature approximately. 

\begin{figure}[htbp]
\begin{center}
 \includegraphics[width=0.35\textwidth]{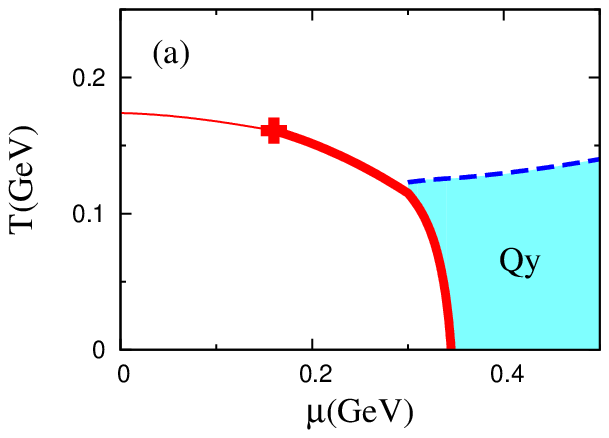} 
 \includegraphics[width=0.35\textwidth]{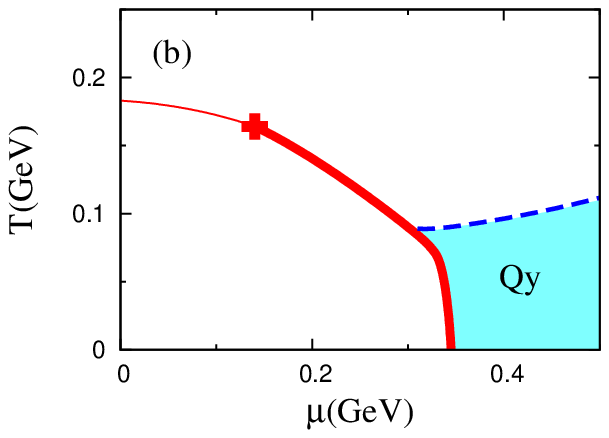} 
\end{center}
\caption{Phase diagram predicted by the EPNJL model with (a) 
constant $T_0$ and (b) $\mu$-dependent $T_0$. 
The thick (thin) solid line corresponds to the chiral transition line of 
first-order (crossover), while  
the dashed line represents a line of $\rho_{1}=\rho_{3}$. 
The hatching region labeled by 'Qy' stands for the quarkyonic phase of 
definition 2. 
}
\label{PD2}
\end{figure}

In Fig.~\ref{PD2}, the quark phase appears out of the chiral transition line 
(the thin and thick solid lines). 
The quark phase is separated into the $\rho_{3}$- and $\rho_{1}$-dominated 
regions by the dashed line, and the $\rho_{3}$-dominated 
($\rho_{1}$-dominated) region is located below (above) the border, 
as expected. 
At $\mu > M$, the $\rho_{3}$-dominated region is much wider than 
the chirally-symmetric confinement region shown in Fig.~\ref{PD}. 
If the quarkyonic phase is defined by the $\rho_{3}$-dominated region, 
the phase appears at small $T$ and $\mu > M$, independently of $\Phi$. 

Now we summarize two definitions of the quarkyonic phase. 
\begin{enumerate}
	\item Phase of $n_{\rm q} \neq 0$ and $\Phi < 0.5$.
    \item Phase of $\rho_{3} > \rho_{1}, \rho_{2}$.
\end{enumerate}
The quarkyonic phase of definition 1 (2) is plotted 
by a region labeled by 'Qy' in 
Fig.~\ref{PD} (Fig.~\ref{PD2}). 
For $\mu > M$, the quarkyonic phase of definition 2 is wider 
than that of definition 1. 
Under definition 2, the region of the quarkyonic phase is independent of 
$\Phi$, and also not so sensitive 
to the choice of constant $T_0$ or $\mu$-dependent $T_0$, as shown by 
the comparison of panels (a) and (b) in Fig.~\ref{PD2}. 

Hereafter, we take definition 1 commonly used. 
We will show in subsections 
\ref{hadronic effect on phase diagram} and 
\ref{Effect of hadronic mass change} that 
the quarkyonic phase of definition 1 survives even after 
hadronic degrees of freedom are taken into account. Hence, so does 
the quarkyonic phase of definition 2.

\subsection{Hadronic effect on phase diagram}
\label{hadronic effect on phase diagram}

At small $T$ and $\mu$, it is natural to think that QCD dynamics is governed by hadronic (mesonic and baryonic) modes than quark modes. 
A simplest way of treating these modes is the free-hadron-gas (FHG) 
approximation. 
Under this approximation, the thermodynamic potential is just a sum 
of the PNJL thermodynamic potential $\Omega_{\rm qrk+glu}$, 
the free-meson thermodynamic potential $\Omega_{\rm msn}$ and 
the free-baryon one $\Omega_{\rm bryn}$: 
\begin{align}
\Omega &= \Omega_{\rm qrk+glu}+\Omega_{\rm msn}+\Omega_{\rm bryn},
\label{Omega_H}\\
\Omega_{\rm msn} &= \sum_{m=\pi,\sigma}
T\int\frac{d^3{\bf q}}{(2\pi)^3}\ln\left(1-e^{-\beta E_m}\right),
\label{Omega-msn} \\
\Omega_{\rm bryn} &= 4T\int\frac{d^3{\bf q}}{(2\pi)^3}
\ln\left(1+e^{-\beta E_b^{-}}\right)
\left(1+e^{-\beta E_b^{+}}\right) ,
\label{Omega-bryn} 
\end{align}
where $E_m=\sqrt{{\bf p}^2+m_m^2}$ and $E_b^\pm=\sqrt{{\bf p}^2+m_b^2} 
\pm 3\mu$ for meson and baryon masses $m_m$ and $m_b$. 
Here, we consider proton and neutron as baryons and $\pi$ and 
$\sigma$ as mesons.

It is well known that hadron has a finite volume. 
Once hadrons are generated 
in the thermodynamic system, some part of the volume of the system is occupied 
by generated hadrons. Owing to this volume-exclusion effect, 
the hadron generation is suppressed at high $T$ and/or high $\mu$ and hence 
the hadronic degrees of freedom do not contribute to EOS there. 
This is the property that EOS should satisfy at high $T$ and/or high $\mu$, 
since hadrons are expected to disappear there. 
For this reason, the volume-exclusion  
effect has often been introduced to EOS. 
The volume-exclusion is described in the canonical ensemble with volume $V$ 
and hadron number $N$ by $\tilde{V}=V-v N$ for 
the excluded volume $v=\frac{4}{3}\pi r^3$ with radius $r$. 
It is possible to convert the definition in the canonical ensemble 
to that in the grand-canonical ensemble by the fugacity and the Laplace 
transform~\cite{Steinheimer,Rischke}. 
In this formulation of Refs.~\cite{Steinheimer,Rischke}, 
the total pressure ${P}^{\rm excl}$ with the volume exclusion effect is 
described by the total pressure $P$  for point-particles as 
\begin{align}
P^{\rm excl} (T, \mu_i)= P (T, \tilde{\mu}_i) ,  
\end{align}
where the $i$ stand for quark, antiquark, meson, baryon and antibaryon. 
In $P$, the chemical potential $\mu_i$ of $i$th particle species is 
replaced by the modified chemical potential 
\begin{align}
\tilde{\mu}_i = \mu_i - v_i P^{\rm excl}
\end{align}
with the excluded volume $v_i=\frac{4}{3}\pi r_i^3$ of radius $r_i$ and 
$\{\mu_i\}=(\mu, -\mu, 0, 3\mu, -3\mu)$ 
for quark, antiquark, meson, baryon and antibaryon, respectively.   
Thus, the volume-exclusion effect can be taken into account simply 
by replacing $\mu_i$ by $\tilde{\mu}_i$ in $P$. 
For simplicity, all hadrons are assumed to have the same volume. 
For $r_i$, we take two cases; the charge radius of nucleon, 0.8~fm, 
and the radius of the repulsive core in the nuclear force, 0.5~fm. 
The number density $n_i$ and the entropy density $s_i$ for the $i$th 
particle species with the volume exclusion effect are 
obtained from the thermodynamic consistency by 
\begin{align}
n_i \equiv \Big(\frac{\partial P^{\rm excl}}{\partial \mu_i}\Big)_{T} 
= f \tilde{n}_i,~~~
s \equiv \Big(\frac{\partial P^{\rm excl}}{\partial T}\Big)_{{\rm all}~\mu_i} 
= \sum_i f\tilde{s}_i,~~~
\end{align}
with 
\bea
f = \frac{1}{1+\sum_i v_i \tilde{n}_i},
\eea
where $\tilde{n}_i$ and $\tilde{s}_i$ are the number and the entropy 
density for point particles 
with the modified chemical potential $\tilde{\mu}_i$. 
This model is called the PNJL+FHG model in this paper. 
In the PNJL+FHG model, the net quark-number density $\rho$ is obtained by 
$
  \rho=n_{\rm q} - {\bar n}_{\rm q} + 3n_{\rm b} - 3{\bar n}_{\rm b} 
$
for the baryon- and antibaryon-number densities 
$n_{\rm b}$ and ${\bar n}_{\rm b}$.

In this subsection, we assume that hadrons keep masses at vacuum even for 
the case of finite $T$ and $\mu$; namely, 
$m_{\pi}=139$~MeV for pion, $m_{\sigma}=680$~MeV for sigma meson and 
$m_{\rm p}=m_{\rm n}=940$~MeV for proton and neutron.  
This approximation is quantitatively valid at $T \lesssim 0.9 T_{\rm c}$, 
but even at $T \gtrsim 0.9 T_{\rm c}$ it is acceptable for qualitative 
discussion, as shown later in subsection \ref{Effect of hadronic mass change}.

\begin{figure}[htbp]
\begin{center}
 \includegraphics[width=0.35\textwidth]{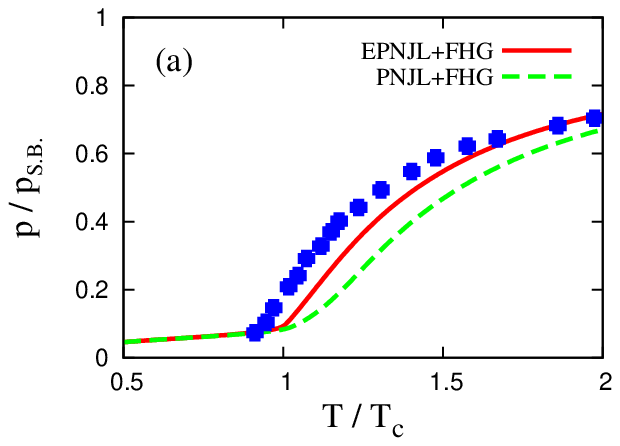} 
 \includegraphics[width=0.35\textwidth]{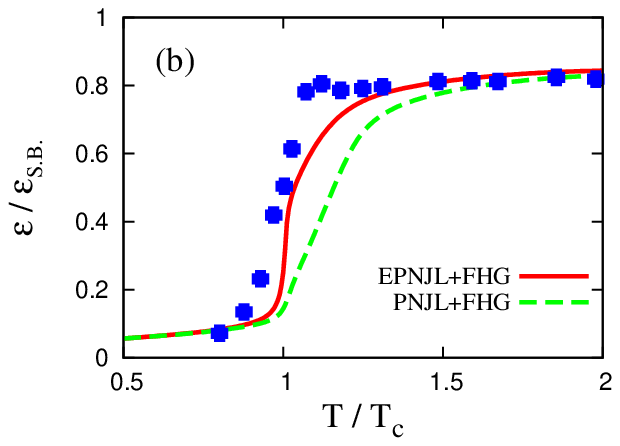} 
 \includegraphics[width=0.35\textwidth]{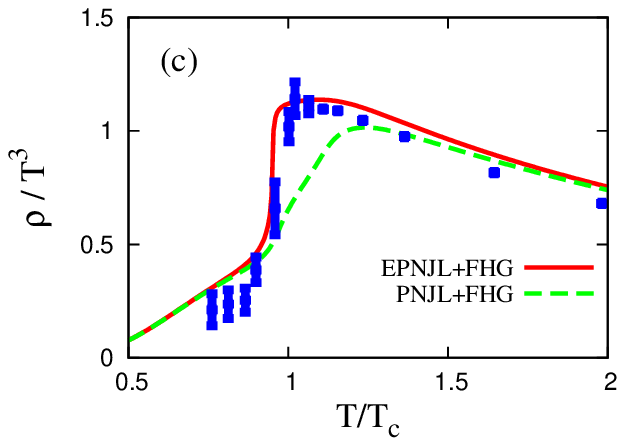} 
\end{center}
\caption{$T$ dependence of (a) the pressure at $\mu=0$, 
(b) the energy density at $\mu=0$ and (c) the net quark-number density 
at $\mu/T_{\rm c}=0.8$. 
The solid (dashed) lines represent results of the EPNJL+FHG (PNJL+FHG) 
model with the volume exclusion of $r=0.8$~fm. 
See Fig.~\ref{PEN1} for other information. 
}
\label{PEN2}
\end{figure}

Figure~\ref{PEN2} represents the pressure, the energy density and 
the quark number density obtained by the PNJL+FHG and EPNJL+FHG models 
with constant hadron masses and the volume-exclusion of $r=0.8$~fm. 
Comparing this figure with Fig.~\ref{PEN1}, one can see for $T < T_{\rm c}$ 
that the PNJL+FHG (EPNJL+FHG) models give 
better agreement with LQCD data than the PNJL (EPNJL) models. 
For $T > T_{\rm c}$, furthermore, 
the EPNJL+FHG model (solid line) yields a better coincidence with LQCD data 
than the PNJL+FHG result (dashed line). 
Thus, the EPNJL+FHG model well describes $T$ dependence of LQCD data 
for all $T$ up to $2 T_{\rm c}$. 

\begin{figure}[htbp]
\begin{center}
 \includegraphics[width=0.35\textwidth]{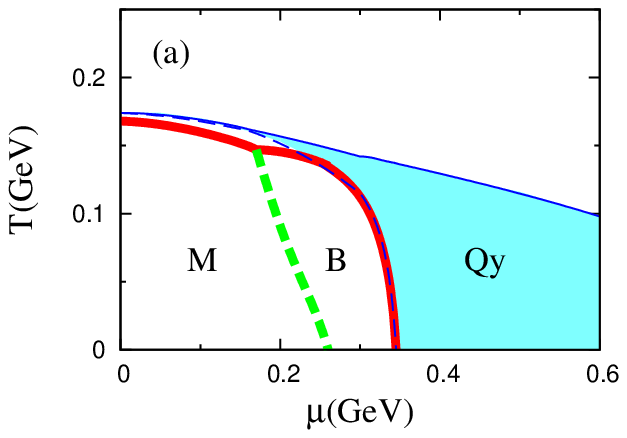} 
 \includegraphics[width=0.35\textwidth]{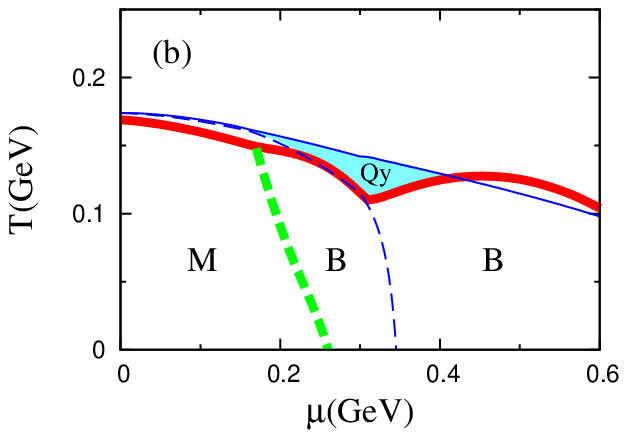} 
\end{center}
\caption{
Dominant degree of freedom in $\mu$-$T$ plane. 
The quark-dominated region is out of the thick-solid line, while 
the meson- and baryon-dominated regions are located inside 
the thick-solid line. 
The meson- and baryon-dominated regions are separated 
by the thick-dashed line, and they are labeled by 'M' and 'B', respectively. 
The hatching area labeled by 'Qy' stands for the quarkyonic phase of 
definition 1. 
The calculation is done by the EPNJL+FHG model 
with (a) $r=0.8$fm and (b) $r=0.5$~fm. 
The thin-solid and thin-dashed lines correspond to 
the confinement and chiral transition lines, respectively. 
}
\label{PD-Mc}
\end{figure}

Figure~\ref{PD-Mc} shows which mode is the main component of EOS 
in $T-\mu$ plane. 
The meson-, baryon- and quark-dominated regions are classified 
by comparing meson, baryon and quark degrees of freedom with each other; 
for example, the meson-dominated region is defined by the condition of 
$n_{\rm m} > n_{\rm b}+{\bar n}_{\rm b}$ and 
$n_{\rm m} > n_{\rm q}+{\bar n}_{\rm q}$, 
where $n_{\rm m}$ is the meson-number density. 
The meson- and the baryon-dominated region are located inside 
the thick-solid line  and they are separated 
by the thick-dashed line and labeled by 'M' and 'B', 
respectively. The results are obtained by the EPNJL+FHG model with 
$r=0.8$~fm in panel (a) and $r=0.5$~fm in panel (b). 
The meson-dominated region is located at small $T$ and $\mu$, 
and the baryon-dominated region is at low $T$ and middle $\mu$. 
The baryon-dominated region is expanded by decreasing $r$, i.e., weaker 
volume-exclusion.

In Fig.~\ref{PD-Mc}, the quark-dominated region is located 
out of the thick-solid line. The thin-solid and thin-dashed lines 
represent the confinement and chiral transition lines, respectively. 
The intersection between the quark-dominated region and the area 
between the thin-solid and thin-dashed lines is the quarkyonic phase of 
definition 1; this is shown by the hatching area with label 'Qy'. 
For stronger volume exclusion of $r=0.8$~fm in panel (a), 
the quarkyonic phase is located at $\mu \gtrsim 340$~MeV 
and $T \lesssim 100$~MeV. 
For weaker volume exclusion of $r=0.5$~fm in panel (b), 
the quarkyonic phase shifts to higher $\mu$, but an island of 
the phase remains around $\mu=340$~MeV and $T=120$~MeV.

\subsection{Effect of $T$- and $\mu$-dependent hadron mass on phase diagram}
\label{Effect of hadronic mass change}
In subsection \ref{hadronic effect on phase diagram}, we considered 
hadron degrees of freedom by using the free gas approximation 
with constant mass. However, hadronic masses are changed with $T$ and $\mu$. 
In this subsection, the $T$ and $\mu$ dependences are taken into account 
in a simple manner.

Meson modes are quantum fluctuations around the mean field and 
can then be calculated with the random phase approximation (RPA). 
Up to order $1/N_{\rm c}$, 
the thermodynamic potential is obtained by~\cite{Zhuang} 
\begin{eqnarray}
\Omega=\Omega_{\rm MF}+\Omega_{\rm RPA}, 
\end{eqnarray}
where $\Omega_{\rm MF}$ is the mean-field part shown in \eqref{eq:E12} 
and $\Omega_{\rm RPA}$ is the mesonic-fluctuation part described by 
the ring diagram~\cite{Zhuang}: 
\begin{eqnarray}
\Omega_{\rm RPA}=\frac{T}{2}\sum_n\int \frac{d^3{\bf q}}{(2\pi)^3}
\ln\det[1-G_{\rm s}\Pi(q)] 
\label{Omegafl}
\end{eqnarray}
with the mesonic polarization bubbles 
\begin{eqnarray}
\Pi_{jk}(q)&=&-T\sum_n\int \frac{d^3{\bf q}}{(2\pi)^3}
{\rm Tr}[\Gamma^*_j{\cal S}(p+q)\Gamma_k{\cal S}(p)]~~~~~
\end{eqnarray}
for $j,~k=\sigma,~\pi_+,~\pi_-,~\pi_0$, 
where Tr is the trace in color, flavor and Dirac indices. 
${\cal S}$ is the quark propagator in the mean-field approximation, 
$
{\cal S}(p)=\left(\gamma_{\nu}(p-A)^{\nu}+M-\mu\gamma_0\right)^{-1}.
$
The meson vertex $\Gamma_k$ depends on meson taken; precisely, 
$\Gamma_{\sigma}=1,~\Gamma_{\pi_+}=i\tau_+\gamma_5,
~\Gamma_{\pi_-}=i\tau_-\gamma_5,~\Gamma_{\pi_0}=i\tau_3\gamma_5$. 

Since it is difficult to calculate the dynamical mesonic fluctuations 
(\ref{Omegafl}) exactly, we then make the pole approximation, that is, 
$\Omega_{\rm RPA}$ is approximated into $\Omega_{\rm msn}$ of 
\eqref{Omega-msn} but with $m_{m}$ replaced by the pole mass $m_j$ 
that satisfies  
\begin{eqnarray}
\det[1-G_{\rm s}\Pi(q_0=m_j,{\bf q=0})]=0. 
\end{eqnarray}

\begin{figure}[htbp]
\begin{center}
 \includegraphics[width=0.35\textwidth]{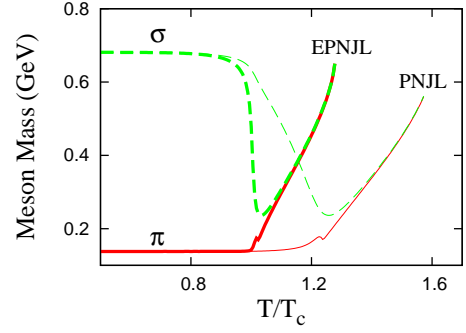} 
\end{center}
\caption{Meson masses as a function of $T$ at $\mu=0$. 
Solid (dashed) lines stand for pi (sigma) meson masses. 
The thick (thin) lines correspond to the EPNJL (PNJL) cases. }
\label{MesonMass}
\end{figure}

Figure~\ref{MesonMass} shows $T$ dependence of $\pi$- and $\sigma$-meson 
masses by solid and dashed lines, respectively. 
The thick and thin lines correspond to results of the EPNJL and PNJL models, 
respectively. 
Since the chiral transition is sharp in the EPNJL model compared with 
in the PNJL model, $T$ dependence of $\sigma$-meson mass is also changed 
rapidly in the EPNJL model.
In the chiral-symmetry broken phase at small $T$ and $\mu$, 
$\pi$-meson masses are small, 
so that $\Omega$ is dominated by $\Omega_{\rm msn}$ in its $\pi$-meson part. 
In the chirally-symmetric phase at high $T$ and/or high $\mu$, 
on the contrary, $\pi$- and $\sigma$-meson 
masses are getting large and hence the mesonic contribution to $\Omega$ 
becomes small.

The baryon mass is hard to obtain even with the RPA approximation. 
In Ref.~\cite{Bentz}, the baryon mass is calculated in the NJL model by 
regarding baryon as a bound state of a quark and a diquark. 
The baryon mass $m_{b}$ is approximately described by 
the constituent quark mass $M$ as  
\begin{eqnarray}
m_{b}\approx 2.24M+0.18~{\rm GeV}. 
\label{BEmass}
\end{eqnarray}
Here, we assume that $\Omega_{\rm bryn}$ of \eqref{Omega-bryn} has 
baryon masses $m_{b}$ of \eqref{BEmass}.

\begin{figure}[htbp]
\begin{center}
 \includegraphics[width=0.35\textwidth]{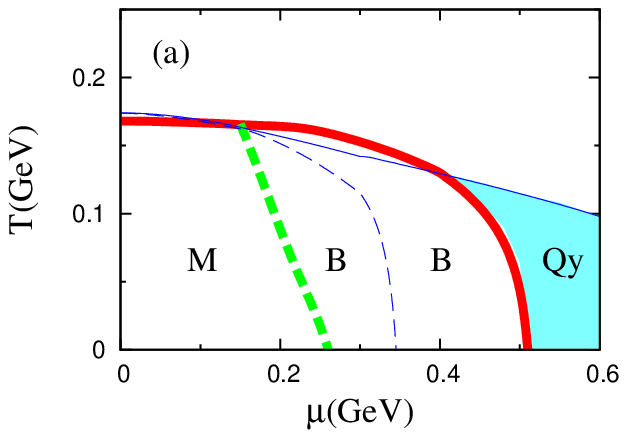} 
 \includegraphics[width=0.35\textwidth]{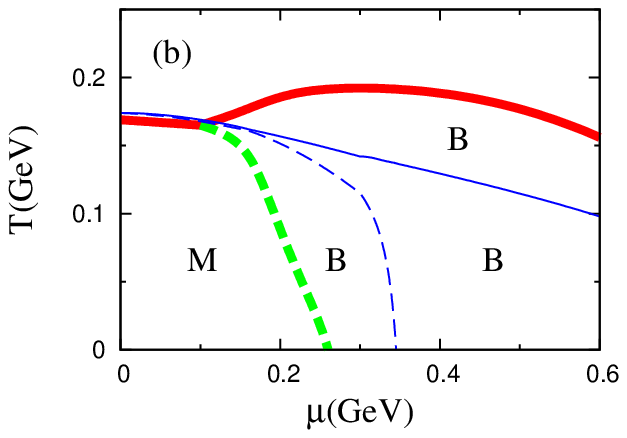} 
\end{center}
\caption{Dominant degree of freedom in $\mu$-$T$ plane.
The calculation is done by the EPNJL+FHG model with 
$T$- and $\mu$-dependent $m_{m}$ and $m_{b}$ and the volume-exclusion effect. 
Here, 
$r=0.8$~fm in panel (a) and $r=0.5$~fm in panel (b). 
In panel (b), the quarkyonic phase emerges at $\mu$ much higher than 
600~MeV. See Fig.~\ref{PD-Mc} for other information. 
}
\label{PD-Mv}
\end{figure}

Now, we have $\Omega$ with $T$- and $\mu$-dependent $m_{m}$ and $m_{b}$. 
Figure~\ref{PD-Mv} is the same figure as Fig.~\ref{PD-Mc}, but 
$T$- and $\mu$-dependence of hadron masses are taken into account 
in Fig.~\ref{PD-Mv}. The variation of hadron masses does not 
change qualitatively locations of baryon-, meson-, quark-dominated regions 
and also a location of the quarkyonic phase; note that in panel (b) 
the quarkyonic phase emerges at $\mu$ much higher than 600~MeV. 
More precisely, the hadron-mass variation 
shifts the boundary between baryon- and quark-dominated regions 
to higher $T$, because of baryon-mass suppression near the boundary.

\section{Summary}
\label{Summary}

The equation of state in two-flavor QCD was investigated 
with the EPNJL model and its simple extension. 
The EPNJL result is consistent with LQCD data on EOS 
at $\mu \ge 0$ better than the PNJL model. 
Thus, the EPNJL model is more reliable than the PNJL model. 
After confirming the reliability of the EPNJL model, 
we have predicted the phase diagram in 
the $\mu$-$T$ plane with the EPNJL model. 
The quarkyonic phase survives, even if the correlation between 
the chiral condensate and the Polyakov loop is strong. 
As an extension of the EPNJL model, we have introduced 
hadronic degrees of freedom by using the free-hadron-gas (FHG) model with 
constant or $T$- and $\mu$-dependent hadron masses. 
Here, the volume exclusion effect 
due to the hadron generation is also taken into account in the FHG model. 
The volume exclusion guarantees that the quark degree of freedom is dominant 
in EOS at high $T$ and/or high $\mu$. 
The EPNJL+FHG model improves agreement of the EPNJL model with 
LQCD data particularly at small $T$. 
The quarkyonic phase survives, even if hadron degrees of freedom are 
taken into account. However, the location of the quarkyonic phase 
is sensitive to the strength of the volume exclusion.

\noindent
\begin{acknowledgments}
The authors thank A. Nakamura and K. Nagata for useful discussions. 
H.K. also thanks M. Imachi, H. Yoneyama, H. Aoki and M. Tachibana for useful discussions. 
Y.S. and T.S. are supported by JSPS.
\end{acknowledgments}


\end{document}